\documentstyle[aps,epsfig]{revtex}
\tighten
\begin{document}
\draft
\twocolumn
\preprint{Preprint Numbers: \parbox[t]{45mm}{KSU-CNR-106-00\\
                                             nucl-th/0005015}}

\title{The $\pi$, $K^+$, and $K^0$ electromagnetic form factors}

\author{Pieter Maris and Peter C. Tandy} 
\address{Center for Nuclear Research, Department of Physics,\\ 
         Kent State University, Kent OH 44242}
\date{\today}
\maketitle
\begin{abstract}
The rainbow truncation of the quark Dyson--Schwinger equation is
combined with the ladder Bethe--Salpeter equation for the meson
amplitudes and the dressed quark-photon vertex in a self-consistent
Poincar\'e-invariant study of the pion and kaon electromagnetic form
factors in impulse approximation.  We demonstrate explicitly that the
current is conserved in this approach and that the obtained results are
independent of the momentum partitioning in the Bethe--Salpeter
amplitudes.  With model gluon parameters previously fixed by the
condensate, the pion mass and decay constant, and the kaon mass, the
charge radii and spacelike form factors are found to be in good
agreement with the experimental data.
\end{abstract}

%

\section{Introduction}

The light pseudoscalar mesons play an important role in understanding
low-energy QCD.  They are the lightest observable hadronic bound states
of a quark and an anti-quark, and are the Goldstone bosons associated
with chiral symmetry breaking.  Their static properties such as the mass
and decay constants have been studied extensively~\cite{mpifpi,MR97}.
Dynamic properties and scattering observables are much less understood
theoretically, but therefore not less important to calculate within QCD.
In this respect, the elastic electromagnetic form factors of the pion
and kaon are very interesting: the probe is well understood, there are
accurate data for $F_\pi$ at low $Q^2$ to confront theoretical
calculations with, and the charge radii $r_\pi^2$, $r_{K^+}^2$, and
$r_{K^0}^2$ are experimentally known.  Currently, there are several
experiments at JLab to determine both the pion and the kaon form factor
in the range $0.5 < Q^2 < 3\,{\rm GeV}^2$ to better
accuracy~\cite{cebafka,cebafpi}, which could help to discriminate
between different model calculations.

To calculate these form factors, we use an approach based on the
Dyson--Schwinger equations [DSEs], which form an excellent tool to study
nonperturbative aspects of hadron properties in QCD~\cite{dserev}.  The
approach is consistent with quark and gluon
confinement~\cite{dserev,conf}, generates dynamical chiral symmetry
breaking~\cite{dcsb}, and is Poincar\'e invariant.  It is
straightforward to implement the correct one-loop renormalization group
behavior of QCD~\cite{MR97}, and obtain agreement with perturbation
theory in the perturbative region.  Provided that the relevant Ward
identities are preserved in the truncation of the DSEs, the
corresponding currents are conserved.  Axial current conservation
induces the Goldstone nature of the pions and kaons~\cite{MRT98};
electromagnetic current conservation produces the correct hadronic
charge without fine-tuning.

We obtain the meson Bethe--Salpeter amplitudes [BSAs] and the
quark-photon vertex as the solutions of respectively the homogeneous and
inhomogeneous Bethe--Salpeter equation [BSE] in ladder truncation.  The
required dressed quark propagators are obtained from solutions of the
quark DSE in rainbow truncation.  Non-analytic effects from vector
mesons are automatically taken into account, because these vector $q\bar
q$ bound states appear as poles in the quark-photon vertex
solution~\cite{MT99pion}.  We employ a realistic model for the effective
quark-antiquark coupling that has been shown to reproduce the pion and
kaon masses and decay constants~\cite{MR97} as well as the masses and
decay constants for the vector mesons $\rho$, $\phi$ and K$^\star$ to
within 10\%~\cite{MT99}.  The model parameters are all fixed in previous
work~\cite{MT99} and constrained only by $m_\pi$, $m_K$, $f_\pi$ and
$\langle\bar q q\rangle$.  The produced pion charge radius is within 2\%
of the experimental value~\cite{MT99pion}.  Here, we use the same
approach, without parameter adjustment, to calculate the neutral and
charged kaon form factors and charge radii; we also extend our previous
pion form factor calculations~\cite{MT99pion} to the spacelike
$Q^2$-domain anticipated for future JLab~\cite{cebafpi} data.

In Sec.~II we review the formulation that underlies a description of the
pion and kaon charge form factors within a modeling of QCD through the
DSEs.  Within the impulse approximation, we outline the manner in which
a ladder-rainbow dynamics for the propagators, BSAs and quark-photon
vertex preserves the meson electromagnetic current.  We further indicate
the additional terms needed for current conservation, if one goes beyond
rainbow-ladder truncation for the DSEs.  In Sec.~III we discuss the
details of the model and present our numerical results for the form
factors.  Concluding remarks are given in Sec.~IV.

\section{Pseudoscalar electromagnetic form factors}

The 3-point function describing the coupling of a photon with momentum
$Q$ to a pseudoscalar meson with initial and final momenta \mbox{$P_\pm
= P \pm Q/2$} respectively can be written as the sum of two terms
\begin{eqnarray}
\label{mesonff}
\Lambda^{a\bar{b}}_\nu(P,Q) &=& 
                \hat{Q}^a \, \Lambda^{a\bar{b}a}_\nu(P,Q) 
                + \hat{Q}^{\bar{b}} \, \Lambda^{a\bar{b}\bar{b}}_\nu(P,Q) \,,
\end{eqnarray}
with $\hat{Q}$ the electric charge of the (anti-)quark, $\case{2}{3}$
for the $u$-quark, and $-\case{1}{3}$ for the $d$- and $s$-quarks, and
with $\Lambda^{a\bar{b}a}$ and $\Lambda^{a\bar{b}\bar{b}}$ describing
the coupling of a photon to the quark and anti-quark inside the meson
respectively.  The meson form factor is defined as
\begin{equation}
 \Lambda^{a\bar{b}}_\nu(P,Q) = 
        2\;P_\nu\;F(Q^2) \,,
\end{equation}
and the corresponding charge radius as $r^2 = -6 F'(Q^2)$ at $Q^2=0$.
Analogously, we can define a form factor for each of the two terms on
the RHS of Eq.~(\ref{mesonff})
\begin{equation}
 \Lambda^{a\bar{b}\bar{b}}_\nu(P,Q) = 
        2\;P_\nu\;F_{a\bar{b}\bar{b}}(Q^2) \,.
\end{equation}
Current conservation dictates that each of the form factors
$F_{a\bar{b}\bar{b}}(Q^2)$ and $F_{a\bar{b}a}(Q^2)$ are 1 at $Q^2=0$.

\subsection{Impulse Approximation}

Using dressed quark propagators, bound state BSAs, and the dressed
$qq\gamma$-vertex, form factors can be calculated in impulse
approximation.  We denote by $\Gamma^a_\mu(q,q';Q)$ the quark-photon
vertex describing the coupling of a photon with momentum $Q$ to a quark
with final and initial momenta $q$ and $q' = q - Q$ respectively and
flavor $a$.  With this notation, the vertices in Eq.~(\ref{mesonff})
take the form\footnote{We use Euclidean metric
$\{\gamma_\mu,\gamma_\nu\} = 2\delta_{\mu\nu}$, $\gamma_\mu^\dagger =
\gamma_\mu$ and $a\cdot b = \sum_{i=1}^4 a_i b_i$.}
\begin{eqnarray}
\lefteqn{\Lambda^{a\bar{b}\bar{b}}_\nu(P,Q) = 
        2\,N_c\int^\Lambda\!\!\frac{d^4k}{(2\pi)^4}
        \,{\rm Tr}\big[ S^a(q) \, \Gamma^{a\bar{b}}(q,q_+;P_-) 
} \nonumber \\ &\times&
        S^b(q_+) \, i \Gamma^{b}_\nu(q_+,q_-;Q)\, S^b(q_-) \,
        \bar\Gamma^{a\bar{b}}(q_-,q;P_+) \big] \;, \;\;\;\;
\label{triangle}
\end{eqnarray}
where 
\mbox{$q = k+\case{1}{2}P$}, 
\mbox{$q_\pm = k-\case{1}{2}P \pm \case{1}{2}Q$},  
\mbox{$P_\pm = P \pm \case{1}{2}Q$},  
and analogously for $\Lambda^{a\bar{b}a}_\nu$.  The notation
$\int^\Lambda$ denotes a translationally-invariant regularization of the
integral, with $\Lambda$ the regularization mass-scale, which can be
removed at the end of all calculations by taking the limit $\Lambda \to
\infty$.  $S(q)$ is the dressed quark propagator and 
$\Gamma^{a\bar{b}}(q,q';P)$ is the meson BSA, with $P^2 = -m^2$ the
on-shell meson momentum, and $q$ and $q'=q-P$ the quark and anti-quark
momenta respectively.

Both $S(q)$ and $\Gamma^{a\bar{b}}(q,q';P)$ are solutions of their
respective DSEs
\begin{eqnarray}
\label{gendse}
\lefteqn{ S(p)^{-1} = Z_2\,i\,/\!\!\!p + Z_4\,m(\mu)}
\nonumber\\       
        &+& Z_1 \int^\Lambda \frac{d^4q}{(2\pi)^4} \,g^2 D_{\mu\nu}(p-q) 
        \frac{\lambda^i}{2}\gamma_\mu S(q)\Gamma^i_\nu(q,p) \,,
\end{eqnarray}
and
\begin{equation}
\Gamma^{a\bar{b}}(p,p';Q) = 
        \int^\Lambda\!\!\frac{d^4q}{(2\pi)^4} 
        K(p,q;Q) \,\chi^{a\bar{b}}(q,q';Q) \,,
\label{homBSE}
\end{equation}
where $D_{\mu\nu}(k)$ is the renormalized dressed-gluon propagator,
$\Gamma^i_\nu(q,p)$ is the renormalized dressed quark-gluon vertex, $K$
is the renormalized $\bar q q$ scattering kernel that is irreducible
with respect to a pair of $\bar q q$ lines, and
\mbox{$\chi^{a\bar{b}}(q,q';Q) = S^a(q)
\Gamma^{a\bar{b}}(q,q';Q) S^b(q')$} is the BS wave function.

The solution of Eq.~(\ref{gendse}) is renormalized according to
$S(p)^{-1} = i\,/\!\!\!p + m(\mu)$ at a sufficiently large spacelike
$\mu^2$, with $m(\mu)$ the renormalized quark mass at the scale $\mu$.
In Eq.~(\ref{gendse}), $S$, $\Gamma^i_\mu$ and $m(\mu)$ depend on the
quark flavor, although we have not indicated this explicitly.  The
renormalization constants $Z_2$ and $Z_4$ depend on the renormalization
point and the regularization mass-scale, but not on flavor: in our
analysis we employ a flavor-independent renormalization scheme.

The meson BSAs $\Gamma^{a\bar{b}}(q,q';P)$ are normalized according to
the canonical normalization condition
\begin{eqnarray}
 \lefteqn{P_\mu = N_c \; \frac{\partial}{\partial P_\mu} 
        \int^\Lambda\!\!\frac{d^4q}{(2\pi)^4}\, \Bigg\{  
        {\rm Tr}\Big[\bar\Gamma^{a\bar{b}}(\tilde{q}',\tilde{q};Q) }
\nonumber\\ &&        
        \times \; S^a(q+\eta P)\, 
        \Gamma^{a\bar{b}}(\tilde{q},\tilde{q}';Q)\,
        S^b(q+(\eta-1)P)\Big] +
\nonumber\\ &&
        \int^\Lambda\!\!\!\frac{d^4k}{(2\pi)^4} 
        {\rm Tr}\left[\bar\chi^{a\bar{b}}(\tilde{k}',\tilde{k};Q)\,
        K(\tilde{k},\tilde{q};P)\,
        \chi^{a\bar{b}}(\tilde{q},\tilde{q}';Q)\right]  
        \Bigg\} \,,
\label{gennorm}
\end{eqnarray}
at the mass shell ${P^2=Q^2=-m^2}$, with \mbox{$\tilde{q}=q+\eta Q$},
\mbox{$\tilde{q}'=q+(\eta-1) Q$}, and similarly for $\tilde{k}$ and 
$\tilde{k}'$.  We use the conventions where \mbox{$f_\pi = 92\,{\rm
MeV}$}, and \mbox{$\eta$} describes the momentum partitioning between
the quark and anti-quark.  Note that physical observables should be
independent of this parameter.

For pseudoscalar bound states the BSA is commonly decomposed
into~\cite{MR97}
\begin{eqnarray}
\label{genpion}
\lefteqn{ \Gamma(k+\eta P,k+(\eta-1)P;P) = }
\nonumber \\
        &=& \gamma_5 \big[ i E(k^2;k\cdot P;\eta) + 
        \;/\!\!\!\! P \, F(k^2;k\cdot P;\eta) 
\nonumber \\
        && {} + \,/\!\!\!k \, G(k^2;k\cdot P;\eta) +
        \sigma_{\mu\nu}\,k_\mu P_\nu \,H(k^2;k\cdot P;\eta) \big]\,,
\end{eqnarray}
with the invariant amplitudes $E$, $F$, $G$ and $H$ being Lorentz scalar
functions of $k^2$ and $k\cdot P=k P\,\cos\theta$.  Subsequently, each
invariant amplitude can be expanded in $k\cdot P$ based on Chebyshev
polynomials
\begin{equation}
\label{expansion}
\label{chebmom}
 f(k^2, k\cdot P; P^2) = \sum_{i=0}^\infty 
        U_i(\cos\theta) (k P)^i f_i(k^2;P^2) \,.
\end{equation}
For charge-parity eigenstates, such as the pion, the invariant
amplitudes $E$, $F$, $G$, and $H$ have a well-defined charge-parity if
one chooses $\eta=\frac{1}{2}$.  Therefore, these amplitudes are either
entirely even ($E$, $F$, and $H$) or odd ($G$) in $k\cdot P$, and one
needs only the even (or odd) Chebyshev moments to completely describe
these amplitudes, which makes this a convenient decomposition.

\subsection{The quark-photon vertex}

The quark-photon vertex is the solution of the renormalized
inhomogeneous BSE
\begin{eqnarray}
\Gamma^a_\mu(p_+,p_-;Q) &=& Z_2\; \gamma_\mu +
        \int^\Lambda\!\!\frac{d^4q}{(2\pi)^4} \; K(p,q;Q) 
\nonumber \\ && {}\times
         S^a(q_+)\; \Gamma^a_\mu(q_+,q_-;Q) \; S^a(q_-) \,, 
\label{verBSE}
\end{eqnarray}
with $p_\pm = p \pm \frac{1}{2}Q$ and $q_\pm= q\pm \frac{1}{2}Q$, and
with the same kernel $K$ as the homogeneous BSE for meson bound states.
Because of gauge invariance, it satisfies the Ward--Takahashi identity
[WTI]
\begin{equation}
i\,Q_\mu \,\Gamma^a_{\mu}(p_+,p_-;Q)  = 
                       S_a^{-1}(p_+) - S_a^{-1}(p_-) \,.
\label{wtid}
\end{equation}

Solutions of the homogeneous version of Eq.~(\ref{verBSE}) at discrete
timelike momenta $Q^2$ define vector meson bound states with masses
\mbox{$m_V^2=-Q^2$}.  It follows that $\Gamma^a_\mu(p;Q)$ has
poles at those locations, and behaves like
\begin{equation}
 \Gamma^a_\mu(p_+,p_-;Q) \rightarrow 
 \frac{\Gamma_\mu^{a\bar{a}\,V}(p_+,p_-;Q) f_V m_V}{Q^2 + m_V^2} \; ,
\label{comres}
\end{equation}
in the vicinity of these bound states, where $\Gamma_\mu^{a\bar{a}\,V}$
is the $a\bar{a}$ vector meson BSA, and $f_V$ the electroweak decay
constant~\cite{MT99}.

For the photon coupled to $u$- and $d$-quarks, this results in a
$\rho$-meson pole at $Q^2 = -0.6\,{\rm GeV}^2$.  For the photon coupled
to $s$-quarks, the first pole is located around $Q^2 = -1.0\,{\rm
GeV}^2$ at the $\phi$-mass.  At the level of the ladder approximation,
which is commonly used in practical calculations, there is no width
generated for the vector meson, and the vertex has real poles.  One
would have to incorporate the open $\pi\pi$ channel in the ladder BSE
kernel to produce a vector meson width; for the vertex, this would
generate an imaginary part beyond the threshold for pion production,
$Q^2 < -4\,m_\pi^2$ in the timelike region.

The full vertex $\Gamma^a_\mu$ can be decomposed into 4 longitudinal
components and 8 transverse components.  The longitudinal components do
not contribute to the form factors.  In Ref.~\cite{MT99pion} it was
shown that only 5 of the 8 transverse components are important for the
pion form factor, in the momentum range $-0.3 < Q^2 < 1.0\,{\rm GeV}^2$
the remaining 3 components contribute less than 1\%.  We expect that
this will also be the case for the kaon form factor, and use the Dirac
amplitudes $T_1$ to $T_5$ of Ref.~\cite{MT99pion} only.

\subsection{Charge conservation}

At \mbox{$Q=0$} the quark-photon vertex is completely specified by the
differential Ward identity
\begin{equation}
i \,\Gamma^b_{\mu}(p,p;0) = 
        \frac{\partial}{\partial p_\mu} S_b^{-1}(p) \; .
\label{wid}
\end{equation}
If this is inserted in Eq.~(\ref{triangle}), one finds after a change of
integration variables $k \rightarrow k-\frac{1}{2}P$
\begin{eqnarray}
\lefteqn{\Lambda^{a\bar{b}\bar{b}}_\nu(P,0) = 
        2 P_\mu F_{a\bar{b}\bar{b}}(0) =
        2\,N_c\int^\Lambda\!\!\frac{d^4q}{(2\pi)^4} }
\nonumber \\ &&
        {\rm Tr}\big[ \bar\Gamma^{a\bar{b}}(q',q;P) \, S^a(q) \, 
        \Gamma^{a\bar{b}}(q,q';P) \,
        \frac{\partial S^b(q-P)}{\partial P} \big] 
        \;, \;\;
\end{eqnarray}
with $q'=q-P$.  Comparing this expression with Eq.~(\ref{gennorm}) with
$\eta = 0$, we recognize that the physical result \mbox{$F(Q^2=0)=1$}
follows directly from the canonical normalization condition for
$\Gamma^{a\bar{b}}$ with a BSE kernel $K$ independent of the meson
momentum $P$.  For the ladder truncation of the kernel, which we
consider in our calculation in the next section, this is the case.

With a general momentum partitioning parameter $\eta$, the relation
between the normalization condition and electromagnetic current
conservation is not so obvious.  However, using a different $\eta$ in
loop diagrams (without external quark lines) is equivalent to a shift in
integration variables.  For processes that are not anomalous, loop
integrals are independent of a shift of integration variables, provided
that such a shift is performed consistently, and that all approximations
employed respect Poincar\'e invariance.  In performing such a shift, one
has to take special care of the BSAs.  The vertex function
$\Gamma(q,q';P)$, as function of the incoming and outgoing quark
momenta, does not depend on $\eta$; it is only in commonly used
decompositions in terms of Lorentz invariant amplitudes such as
Eq.~(\ref{genpion}), where $\eta$ becomes relevant.  The amplitudes $E$,
$F$, $G$, and $H$ are scalar functions of $k^2$ and $k\cdot P$, which
{\em do} depend on the choice for $\eta$.  Under a change of $\eta$,
some of the different Dirac structures (e.g. the amplitudes $F$ and $G$)
will mix, as will the Chebyshev moments, $f_i$ in Eq.~(\ref{chebmom}).
Therefore, the results will be independent of the momentum routing in
the loop integrals if and only if all Dirac amplitudes {\em and} their
dependence on $k\cdot P$ are properly taken into account.  Previously it
has been shown that under these conditions the decay constants are
indeed independent of $\eta$~\cite{MR97}.

Use of a bare quark-photon vertex, in combination with dressed
propagators, in Eq.~(\ref{triangle}), clearly violates charge
conservation and leads to $F_\pi(0) \neq 1$.  With the Ball--Chiu
Ansatz~\cite{BC80}, which is commonly used in DSE studies of
electromagnetic interactions~\cite{R96,BRT96,pctrev,MR98,HP99,BRSBF99},
the electromagnetic current is explicitly conserved, $F(Q^2=0)=1$.
However, the behavior of the form factor away from $Q^2=0$ is not
constrained by current conservation, and in the present model, use of
the Ball--Chiu Ansatz leads to a value for $r_\pi^2$ which is about 50\%
too small~\cite{MT99pion}.  With the quark-photon vertex as the solution
of the ladder BSE, together with quark propagators from the rainbow DSE,
we satisfy all constraints from current conservation, and the calculated
value of $r^2_\pi$ is within 5\% of the experimental
value~\cite{MT99pion}.

\subsection{Beyond rainbow-ladder truncation}

\begin{figure}
\centering{\
\epsfig{figure=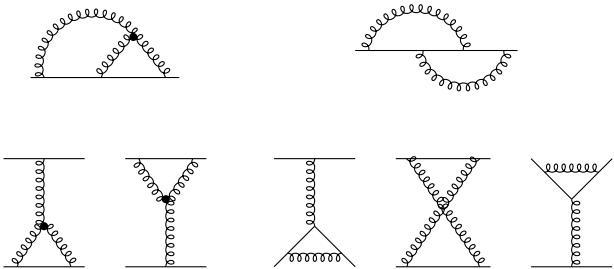, height=4cm} }
\caption{ 
The two leading-order vertex corrections to the rainbow DSE (top) and
the corresponding five diagrams to be added to the ladder BSE kernel
(bottom) for consistency with the relevant WTIs.  The quark and gluon
lines indicate dressed propagators in this and the subsequent figures.
\label{DSEbeyond} }
\end{figure}

If one goes beyond the rainbow-ladder truncation for the DSEs for the
propagators, BSAs and quark-photon vertex, one has to go beyond impulse
approximation for the form factors in order to ensure current
conservation.  For example, one could include higher-order $\alpha_s$
corrections to the rainbow-ladder DSE and BSE kernels, as depicted in
Fig.~\ref{DSEbeyond}.  Following the general procedure developed in
Ref.~\cite{BRS96}, one can show that both the WTI, Eq.~(\ref{wtid}), and
the differential Ward identity, Eq.~(\ref{wid}), are preserved in the
truncation indicated in Fig.~\ref{DSEbeyond}, as is the axial-vector
WTI, which is important for the Goldstone nature of the pions.

\begin{figure}
\centering{\
\epsfig{figure=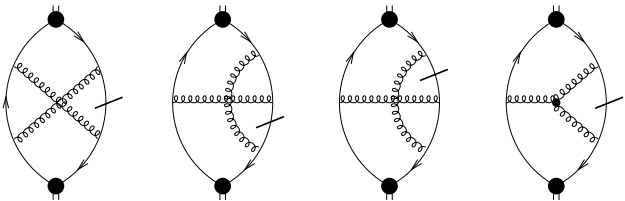, height=3cm} }
\caption{ 
The four diagrams due to the $P$-dependence of the kernel in the
normalization condition, Eq.~(\protect\ref{gennorm}), if one includes
the diagrams of Fig.~\protect\ref{DSEbeyond} into the DSE dynamics, and
chooses all of the meson momentum $P$ to flow through one quark only.
The derivatives with respect to $P$ are marked by slashes.
\label{normbeyond} }
\end{figure}

The resulting BSE kernel $K(q,p;P)$ now becomes dependent on the meson
momentum $P$, which means that the second term of the normalization
condition, Eq.(\ref{gennorm}), is nonzero.  To be specific, with the
choice $\eta = 0$, this introduces the four extra terms in the
normalization condition, diagrammatically depicted in
Fig.~\ref{normbeyond}.  These four additional diagrams can be generated
from the BSE kernel in the bottom part of Fig.~\ref{DSEbeyond} by taking
the derivative with respect to the meson momentum $P$, where $P$ flows
through one quark propagator only.

\begin{figure}
\centering{\
\epsfig{figure=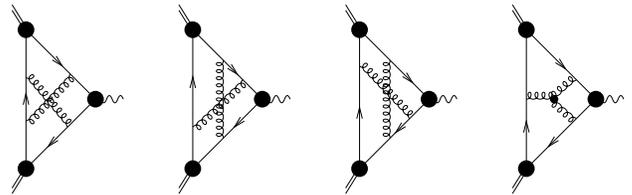, height=3cm} }
\caption{ 
The four additional contributions to the impulse approximation,
Eq.~(\protect\ref{triangle}), in order to ensure current conservation if
one includes the diagrams of Fig.~\protect\ref{DSEbeyond} into the DSE
dynamics.
\label{tribeyond} }
\end{figure}

Since taking the derivative with respect to $P$ is equivalent to the
insertion of a zero-momentum photon according to the differential WTI,
Eq.~(\ref{wtid}), it is obvious which diagrams have to be added to the
impulse approximation to ensure current conservation, see
Fig.~(\ref{tribeyond}).  In the limit $Q\rightarrow 0$ these four
additional diagrams become identical to the four additional diagrams in
Fig.~(\ref{normbeyond}), provided that the vertex satisfies the
differential WTI.  Of course, there are similar contributions to
$\Lambda^{a\bar{b}a}$, which can be identified with terms in the
normalization condition with $\eta = 1$.

Also, simple addition of contributions due to pion and kaon loops to
Eq.~(\ref{triangle}), in combination with a ladder-rainbow truncation
for the DSEs, will generally violate current conservation.  Current
conservation requires a consistent treatment of the kernels for both the
DSE and BSE equations and the approximation for the photon-hadron
coupling.  At present it is not clear how to incorporate meson loops
self-consistently in such an approach, but we expect corrections coming
from such loops to be small in the spacelike region.  In
Ref.~\cite{ABR95} it was demonstrated that the quark core can generate
most of the pion charge radius, and that pion loops contribute less
then 15\% to $r_\pi^2$.  For larger values of $Q^2$ the effect from
meson loops reduces even further, and for \mbox{$Q^2 > 1\,{\rm GeV}^2$}
we expect the contribution of such loops to be negligible.

\section{Model calculations}

For the BSE we use a ladder truncation
\begin{equation}
\label{ourBSEansatz}
        K(p,q;P) \to
        -{\cal G}(k^2)\, D_{\mu\nu}^{\rm free}(k)
        \textstyle{\frac{\lambda^a}{2}}\gamma_\mu \otimes
        \textstyle{\frac{\lambda^a}{2}}\gamma_\nu \,,
\end{equation}
where $D_{\mu\nu}^{\rm free}(k=p-q)$ is the free gluon propagator in
Landau gauge. The resulting BSE is consistent with a rainbow truncation
\mbox{$\Gamma^a_\nu(q,p) \rightarrow \gamma_\nu\lambda^a/2$} for the 
quark DSE, Eq.~(\ref{gendse}), in the sense that the combination
produces vector and axial-vector vertices satisfying the respective
WTIs.  In the axial case, this ensures that in the chiral limit the
ground state pseudoscalar mesons are the massless Goldstone bosons
associated with chiral symmetry breaking~\cite{MR97,MRT98}.  In the
vector case, this ensures electromagnetic current conservation.

The model is completely specified once a form is chosen for the
``effective coupling'' ${\cal G}(k^2)$.  We employ the
Ansatz~\cite{MR97,MT99}
\begin{eqnarray}
\label{gvk2}
\frac{{\cal G}(k^2)}{k^2} &=&
        \frac{4\pi^2\, D \,k^2}{\omega^6} \, {\rm e}^{-k^2/\omega^2}
\nonumber \\  & &              
{}+ \frac{ 4\pi^2\, \gamma_m \; {\cal F}(k^2)}
        {\case{1}{2} \ln\left[\tau + 
        \left(1 + k^2/\Lambda_{\rm QCD}^2\right)^2\right]} \;,
\end{eqnarray}
with \mbox{$\gamma_m=12/(33-2N_f)$} and
\mbox{${\cal F}(s)=(1 - \exp\frac{-s}{4 m_t^2})/s$}.
This Ansatz preserves the one-loop renormalization group behavior of
QCD, and ensures that we reproduce perturbation theory in the
perturbative region.  The first term of Eq.~(\ref{gvk2}) implements the
strong infrared enhancement in the region \mbox{$0 < k^2 < 1\,{\rm
GeV}^2$} which is a phenomenological requirement for sufficient
dynamical chiral symmetry breaking to produce an acceptable strength for
the quark condensate~\cite{HMR98}.  We use
\mbox{$m_t=0.5\,{\rm GeV}$}, \mbox{$\tau={\rm e}^2-1$}, \mbox{$N_f=4$}, 
\mbox{$\Lambda_{\rm QCD} = 0.234\,{\rm GeV}$}, and a renormalization 
point \mbox{$\mu=19\,{\rm GeV}$}, well in the perturbative
region~\cite{MR97,MT99}.  The remaining parameters, \mbox{$\omega =
0.4\,{\rm GeV}$} and \mbox{$D=0.93\,{\rm GeV}^2$}, are fitted to give a
good description of the chiral condensate, $m_{\pi/K}$ and $f_{\pi}$.
The subsequent values for $f_K$ and the masses and decay constants of
the vector mesons $\rho, \phi, K^\star $ are in agreement with the
experimental data~\cite{MT99}, see Table~\ref{sumres}.
\begin{table}
\begin{center}
\caption{\label{sumres}
Overview of the results of the model for the meson masses and decay
constant, adapted from Refs.~\protect\cite{MR97,MT99}.  The experimental
value for the condensate is taken from Ref.~\protect\cite{Lein97}.}
\begin{tabular}{l|ddr}
        & \multicolumn{1}{r}{experiment~\protect\cite{PDG}}  
        & \multicolumn{1}{r}{calculated}  &\\
        & \multicolumn{1}{r}{(estimates)} 
        & \multicolumn{1}{r}{($^\dagger$ fitted)} &\\ \hline
$m^{u=d}_{\mu=1 {\rm GeV}}$ &
        \multicolumn{1}{r}{ 5 - 10 MeV}  & 
        \multicolumn{1}{r}{ 5.5 MeV}     &\\
$m^{s}_{\mu=1 {\rm GeV}}$ &
        \multicolumn{1}{r}{ 100 - 300 MeV} &
        \multicolumn{1}{r}{ 125 MeV   }    &\\ \hline
- $\langle \bar q q \rangle^0_{\mu}$
                & (0.236 GeV)$^3$ & (0.241$^\dagger$)$^3$ &\\
$m_\pi$         &  0.1385 GeV &   0.138$^\dagger$ &\\
$f_\pi$         &  0.0924 GeV &   0.093$^\dagger$ &\\
$m_K$           &  0.496 GeV  &   0.497$^\dagger$ &\\
$f_K$           &  0.113 GeV  &   0.109        &\\ \hline
$m_\rho$        &  0.770 GeV  &   0.742        &\\
$f_\rho$        &  0.216 GeV  &   0.207        &\\
$m_{K^\star}$   &  0.892 GeV  &   0.936        &\\
$f_{K^\star}$   &  0.225 GeV  &   0.241        &\\
$m_\phi$        &  1.020 GeV  &   1.072        &\\
$f_\phi$        &  0.236 GeV  &   0.259        &\\
\end{tabular}
\end{center}
\end{table}
%

\subsection{Results for $u \bar{u} u$, $u \bar{s} u$, and 
$u\bar s \bar s$ form factors}

\begin{figure}
\centering{\
\epsfig{figure=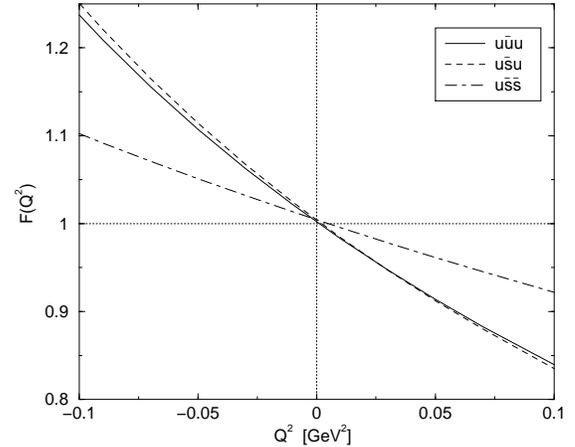, height=6cm} }
\caption{ 
The three independent form factors $F_{u\bar{u}u}$, $F_{u\bar{s}u}$ 
and $F_{u\bar{s}\bar{s}}$.
\label{fig:usssuu} }
\end{figure}

The pion and kaon form factors are given by
\begin{eqnarray}
\label{fpi}
 F_{\pi}(Q^2) &=& \case{2}{3}F_{u\bar{d}u}(Q^2) 
        + \case{1}{3}F_{u\bar{d}\bar{d}}(Q^2)            \,,\\
\label{fKplus}
 F_{K^+}(Q^2) &=& \case{2}{3}F_{u\bar{s}u}(Q^2)  
        + \case{1}{3}F_{u\bar s \bar s}(Q^2)            \,,\\
\label{fK0}
 F_{K^0}(Q^2) &=& -\case{1}{3}F_{d\bar{s}d}(Q^2) 
        + \case{1}{3}F_{d\bar s \bar s}(Q^2)            \,,
\end{eqnarray}
where the quark and anti-quark charges are evident.  We work in the SU(2)
isospin limit, where the strong interaction does not discriminate
between $u$- and $d$-quarks, so for the pion we simply have
$F_{\pi}(Q^2) = F_{u\bar{u}u}(Q^2)$.  Thus there are only three
independent form factors, $F_{u\bar{u}u}(Q^2)$,
$F_{u\bar{s}u}(Q^2)$, and $F_{u\bar s\bar s}(Q^2)$, which are shown in
Fig.~\ref{fig:usssuu}.  Our estimate of the numerical error in these
calculations is less than 1\% for $F_{u\bar{u}u}(Q^2)$, and 2\%
for the other two form factors.

For the pion we use only the leading terms of the expansion of the BSAs
and the quark-photon vertex in $k\cdot P$, Eq.~(\ref{expansion}).
Higher order terms do not change the results more than 1\% in this
momentum regime, although they are needed at larger values of $Q^2$.
For the kaon we have to use more terms in the expansion, even at
$Q^2=0$, to obtain independence from the parameter $\eta$, and to ensure
current conservation.  With terms up to order $(k\cdot P)^1$ only, there
is a spread in our results of more than 10\% at $Q^2=0$ (from 0.94 to
1.06) if we change $\eta$ between 0 and 1.  Including the next two terms
reduces that spread to less then 3\%, illustrating that the result of a
loop integral is independent of this unphysical parameter $\eta$,
provided that all relevant Dirac structures and the dependence on
$k\cdot P$ are properly taken into account.

The results for $F_{u \bar{u}u}$ and $F_{u\bar{s}u}$ are remarkably
close to each other, indicating that the flavor of the spectator quark
matters very little.  Within our numerical errors, they are almost
indistinguishable on the $Q^2$ domain shown.  There is a slight
difference in the slope of these form factors:
\mbox{$r_{u \bar{u}u}^2 = 0.45\,{\rm fm}^2$} versus 
\mbox{$r_{u \bar{s}u}^2 = 0.47\,{\rm fm}^2$}.  These results 
are in good agreement with the pion charge radius, \mbox{$r_\pi^2 =
0.46\,{\rm fm}^2$}, obtained in Ref.~\cite{MT99pion} using all eight
Dirac amplitudes of the quark-photon vertex.

The result for $F_{u\bar{s}\bar{s}}$ is quite different in that it has 
a significantly smaller slope characterized by a radius parameter
\mbox{$r_{u \bar{s}\bar{s}}^2 = 0.21\,{\rm fm}^2$}.  This is due to the 
larger mass of the strange quark, and as a consequence the neutral kaon
charge radius $r_{K^0}^2$ will be negative.  A similar effect was
observed for the neutron form factor, where the heavier mass of the
$0^+(ud)$-diquark compared to the $d$ quark mass leads to a negative
charge radius~\cite{BRSBF99}.  Our result is also consistent with the
qualitative aspects of the vector meson dominance [VMD] picture: the
lowest-mass bound state pole in the $ss\gamma$-vertex is the $\phi$, at
$Q^2 = -1.0\,{\rm GeV}^2$, which is significantly further from the
photon point than is the $\rho$ pole in the $uu\gamma$-vertex at $Q^2 =
-0.6\,{\rm GeV}^2$.  This observation, as well as the difference between
$r_{u\bar{s}\bar{s}}^2$ and $r_{u \bar{s}u}^2$, is consistent with the
larger mass of the strange quark.

\subsection{Results for the meson form factors}

\begin{figure}
\centering{\
\epsfig{figure=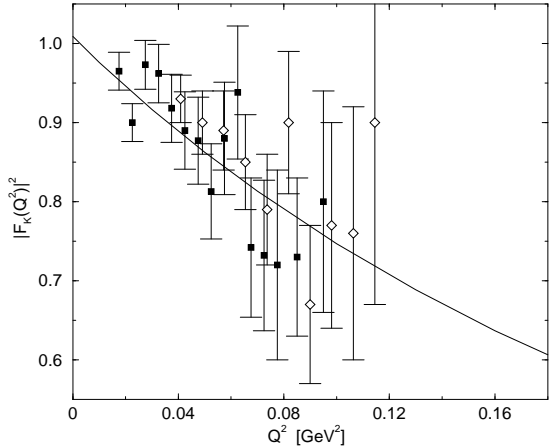, height=6cm} }
\caption{ 
The calculated $K^+$ form factor compared to the data from
Refs.~\protect\cite{Dally80} (open diamonds) and \protect\cite{A86K}
(solid squares).  Within numerical errors, $|F_{K^+}(0)|^2 = 1$.
\label{fig:Kplus} }
\end{figure}

The results in this model for the pion form factor at low $Q^2$, in
particular the pion charge radius, were presented
previously~\cite{MT99pion}.  The obtained charge radii for the kaon are
presented in Table~\ref{resrK}, and are in reasonable agreement with the
experimental data, without any readjustment of the model.  In
Fig.~\ref{fig:Kplus} we show our result for the charged kaon, which is
in good agreement with the available data.
\begin{table}
\begin{center}
\caption{\label{resrK}
Our results for the charge radii, compared with the experimental values
given in Refs.~\protect\cite{A86,A86K,M78}.}
\begin{tabular}{l|cdr}
charge radii & experiment              & calculated &\\ \hline
$r_\pi^2$    & $ 0.44 \pm 0.01 $ fm$^2$&  0.45  fm$^2$ & \\
$r_{K^+}^2$  & $ 0.34 \pm 0.05 $ fm$^2$&  0.38  fm$^2$ & \\
$r_{K^0}^2$  & $-0.054\pm 0.026$ fm$^2$& $-$0.086 fm$^2$ &
\end{tabular}
\end{center}
\end{table}

\begin{figure}
\centering{\
\epsfig{figure=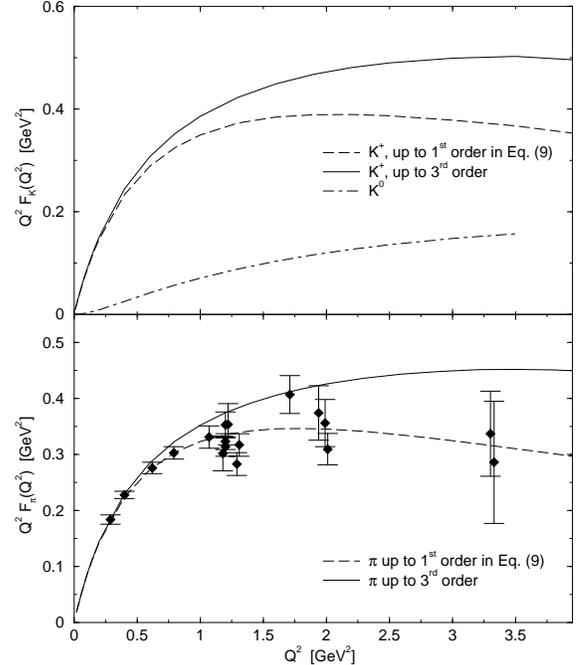, height=9cm} }
\caption{ 
$Q^2$ times the kaon form factors (top) and pion form factor (bottom).
The pion data are from Ref.~\protect\cite{B78}.
\label{fig:Q2FK} }
\end{figure}

Finally, in Fig.~\ref{fig:Q2FK} we present $Q^2 F(Q^2)$ for $\pi$ and
K$^{0,\pm}$ for a larger $Q^2$ range to anticipate data that may be
forthcoming from experiments at JLab~\cite{cebafka,cebafpi} and possibly
other facilities in the future.  In this momentum range, even for
$F_\pi(Q^2)$ the dependence on $k\cdot P$ becomes important, and terms
up to $(k\cdot P)^3$ in Eq.~(\ref{expansion}) are required to produce a
converged result at $Q^2 = 1 \sim 3\,{\rm GeV}^2$.  Higher-order terms
do not change the results by more than 1\% in this momentum range.  Our
estimate is that the net numerical accuracy for $F_{u\bar{u}u}$,
$F_{u\bar{s}u}$, and $F_{u\bar{s}\bar{s}}$ is about 2-3\% at these
values of $Q^2$.  This translates to a similar level of accuracy for
$F_\pi$ and $F_{K^+}$, and to a somewhat larger relative error, about
5\%, for $F_{K^0}$, which is the difference of $F_{u\bar{s}\bar{s}}$ and
$F_{u\bar{s}u}$.  At $Q^2 > 3\,{\rm GeV}^2$, higher-order Chebyshev
moments may be necessary, but current numerical methods prevent their
accurate determination at large $Q^2$.

Over the entire spacelike momentum range considered, \mbox{$F_\pi(Q^2) <
F_{K^+}(Q^2)$}, and $Q^2 F(Q^2)$ rises with $Q^2$ until $Q^2 = 3\,{\rm
GeV}^2$ for all three form factors.  In this momentum range our results
for both the pion and the $K^+$ form factor can be fitted quite well by
a simple monopole $m^2 / (Q^2 + m^2)$, with a mass $m^2=0.53\,{\rm
GeV}^2$ for the pion and $m^2 = 0.61\,{\rm GeV}^2$ for the $K^+$.  A VMD
model, two monopoles for the two form factors $F_{u\bar{s}u}$ and
$F_{u\bar{s}\bar{s}}$ in Eqs.~(\ref{fKplus}) and (\ref{fK0}) with the
physical $\rho$ and $\phi$ masses respectively, does not reproduce our
results for the kaon form factors very well.  For example, at
\mbox{$Q^2 = 1\,{\rm GeV}^2$}, VMD overshoots our $F_{K^+}$ calculation 
by almost 10\%, whereas the monopole fit is within 2\% of our result.
This difference between VMD and our calculations grows with $Q^2$.

Above $Q^2 \sim 3.5\,{\rm GeV}^2$ the monopole fits begin to deviate
significantly from our results and $Q^2 F(Q^2)$ starts to decrease.  In
a more realistic model, that takes meson loop corrections into account
self-consistently, it is very well conceivable that this turn-over
happens at somewhat lower values of $Q^2$: meson loops are expected to
contribute up to 15\% to $r_\pi^2$~\cite{ABR95}, but their
contribution to the form factor decreases rapidly with increasing
spacelike momenta.  In the presence of meson loop corrections the
contribution to the form factor from the impulse approximation has to be
smaller than in our calculation in order to maintain agreement with the
low-$Q^2$ data.  Therefore it is not unlikely that at intermediate
momenta in the present approach we overestimate the form factors, which
may explain the difference between the data points at $Q^2 = 3.3\,{\rm
GeV}^2$ and our calculated results.  More accurate results from JLab, in
combination with realistic model calculations that include meson loop
corrections self-consistently, may be able to resolve this question.

At asymptotically large $Q^2$, factorized pQCD~\cite{BPold} predicts
that the form factor behaves like \mbox{$Q^2 F(Q^2) \to c$}, with
\mbox{$c = 16\pi f_\pi^2\alpha_s(Q^2)$}.  Since our truncation and the
Ansatz, Eq.~(\ref{gvk2}), is constructed so as to preserve asymptotic
freedom, we are guaranteed to recover the leading power-law asymptotic
behavior.  An explicit verification of this behavior, and calculation of
the constant $c$, is not readily available within our present framework
since numerical accuracy at large $Q^2$ is problematic.  However, it is
clear from our results that at $Q^2 \sim 4\,{\rm GeV}^2$ the form factor
has not yet reached its asymptotic value, and it is unlikely that
experiments can access the true asymptotic region in the near future.
In simplified models such as that of Ref.~\cite{MR98} however, it is
straightforward to demonstrate that the impulse approximation does
indeed lead to the power-law behavior predicted by pQCD.

\section{Summary}

We calculate the pion and kaon electromagnetic form factors within the
DSE approach.  The method is completely Poincar\'e invariant, and the
only approximation made is a self-consistent truncation of the set of
DSEs, which respects the relevant vector and axial-vector WTIs.  The
employed quark propagators, the meson BSAs, and the quark-photon vertex
are solutions of their DSEs in rainbow-ladder truncation with all
parameters fixed previously by fitting the chiral condensate,
$m_{\pi/K}$ and $f_{\pi}$.  We include all relevant Dirac amplitudes for
the BSAs and their dependence upon $k\cdot P$.  The electromagnetic
current is explicitly conserved in this approach, and there is no
fine-tuning needed to obtain $F_\pi(0) = 1 = F_{K^+}(0)$ and
$F_{K^0}(0)=0$.  We also demonstrate explicitly that our results are
(within numerical accuracy) independent of the momentum partitioning of
the BSAs.  The obtained pion and kaon form factors are in good agreement
with the available data over the entire $Q^2$ range considered, and the
calculated charge radii are within the error bars of their experimental
values.

These charge radii are somewhat larger than those obtained in a previous
study~\cite{BRT96} that was framed in terms of semi-phenomenological
representations for BSAs and confined quark propagators within the
impulse approximation.  The main difference with that work is that here
we use numerical solutions of truncated DSEs for all the elements needed
in Eq.~(\ref{triangle}), and that all our parameters were fixed
previously.  In comparison with theoretical calculations based on other
methods, it is interesting to note that our results are very similar to
those obtained in Ref.~\cite{choiji}, in particular for the neutral
kaon.

At intermediate values of $Q^2$ our calculations are qualitatively
similar to those obtained in both Ref.~\cite{BRT96} and
Ref.~\cite{choiji}.  Up to about \mbox{$Q^2 = 3\,{\rm GeV}^2$}, both
$F_\pi$ and $F_{K^+}$ can be fitted quite well by a monopole form, with
monopole masses of \mbox{$m^2=0.53\,{\rm GeV}^2$} and
\mbox{$m^2 = 0.61\,{\rm GeV}^2$} respectively.  At large $Q^2$ the DSE 
approach does reproduce the pQCD power-law behavior~\cite{BPold}, but
this behavior does not occur until well beyond~\cite{MR98} the $Q^2$
range considered in our present calculations and accessible at current
accelerators.

\acknowledgements 
We acknowledge useful conversations and correspondence with
C.D.~Roberts, D.~Jarecke and S.R.~Cotanch.  This work was funded by the
National Science Foundation under grant No.~PHY97-22429, and benefited
from the resources of the National Energy Research Scientific Computing
Center.

%
%

%
%
%
%
%
%
%
\end{document}